\documentclass[preprints,articles,accept,moreauthors,pdftex]{Definitions/mdpi} 

\firstpage{1} 
\pubvolume{1}
\issuenum{1}
\articlenumber{0}
\pubyear{2021}
\copyrightyear{2020}
\datereceived{} 
\dateaccepted{} 
\datepublished{} 
\hreflink{https://doi.org/} 

\usepackage{bm}

\Title{Optical forces on an oscillating dipole near VO$_2$ phase transition}

\TitleCitation{Optical forces on an oscillating dipole near VO$_2$ phase transition}

\Author{Daniela Szilard $^{1,}$*\orcidA{}, Patrícia P. Abrantes $^{1}$\orcidB{}, Felipe A. Pinheiro $^{1}$\orcidC{}, Felipe S. S. Rosa $^{1}$\orcidD{}, Carlos Farina $^{1}$\orcidE{} and Wilton J. M. Kort-Kamp $^{2}$\orcidF{}}

\AuthorNames{Daniela Szilard, Patrícia P. Abrantes, Felipe A. Pinheiro, Felipe S. S. Rosa, Carlos Farina and Wilton J. M. Kort-Kamp}

\AuthorCitation{Szilard, D.; Abrantes, P.P.; Pinheiro, F.A.; Rosa, F.S.S.; Farina, C.; Kort-Kamp, W.J.M.}

\address{%
$^{1}$ \quad Instituto de Física, Universidade Federal do Rio de Janeiro, 21941-972, RJ, Brazil; patricia@pos.if.ufrj.br (P.P.A.); fpinheiro@if.ufrj.br (F.A.P.); frosa@if.ufrj.br (F.S.S.R.); farina@if.ufrj.br (C.F.)\\
$^{2}$ \quad Theoretical Division, Los Alamos National Laboratory, MS B262,  Los Alamos, New Mexico 87545, USA; kortkamp@lanl.gov}

\corres{Correspondence: daniela@if.ufrj.br}

\abstract{We investigate optical forces on oscillating dipoles close to a phase-change vanadium dioxide (VO$_2$) film, which exhibits a metal-insulator transition around $340$ K and low thermal hysteresis. This configuration is related to one composed of an excited two-level quantum emitter and we employ a classical description to capture important aspects of the radiation-matter interaction. We consider both electric and magnetic dipoles for two different configurations, namely, with the dipole moments parallel and perpendicular to the VO$_2$ film. By using Bruggeman theory to describe the effective optical response of the material, we show that, in the near-field regime, the force on the dipoles can change from attractive to repulsive just by heating the film for a selected frequency range. We demonstrate that the thermal hysteresis present in the VO$_2$ transition clearly shows up in the behavior of the optical forces, setting the grounds for alternative approaches to control light-matter interactions using phase-change materials.}


\usepackage{ulem} 

\begin{document}


\section{Introduction}


Optical forces play a pivotal role in photonics with many applications. As interesting examples, we can mention radiation pressure forces \cite{Ashkin1970,Ashkin1971}, forces in optical tweezers \cite{Grier2003,Padgett2011,Dholakia2011,Fortuno-EnghetaPRL-2014}, nanostructures \cite{GarciaAbajo2007,Juan2011}, waveguides \cite{Yang2009,Zayats2014}, as well as interdisciplinary applications in biology~\cite{xin2020optical} and atomic physics~\cite{Ashkin1970,Bagnato1987,Phillips1998,Christodoulides2008,Dholakia2011}. Hence, the possibility of harnessing light-matter interactions to tailor and control optical forces at the nanoscale is a sought-after goal in nanophotonics. In particular, the feasibility of switching on and off the repulsion between particles and surfaces in micro- and nanomechanical devices can lead to new functionalities such as the levitation of objects from surfaces to eliminate undesired adhesion and stiction of nanomechanical components \cite{rodriguez2014electric,rodriguez2016repulsion}.

Recent advances in plasmonics and metamaterials allow for the development of new material plataforms to tune optical forces at increasingly smaller scales. Remarkable examples are phase-change materials~\cite{jeong2020dynamic}, such as transition metal dichalcogenides~\cite{van2020exciton}, transparent conductive oxides~\cite{howes2018dynamic}, and liquid crystals~\cite{komar2018dynamic,li2019phase}, which have been integrated into metasurfaces and metadevices to allow for external control of their functionalities. Using this strategy one can either progressively tune or abruptly switch the structural and/or optical properties of metadevices by externally varying an applied voltage~\cite{van2020exciton}, electric current~\cite{berto2019tunable}, and incident light intensity~\cite{shcherbakov2015ultrafast}.  Amid the phase-change materials for photonic applications, vanadium dioxide (VO$_2$) may be singled out for its low thermal hysteresis, for exhibiting a metal-insulator transition (MIT) at low temperature (around $340$ K) over a broad frequency range, and for its high refractive-index contrast in the visible range~\cite{pergament2013oxide,cueff2015dynamic,cavalleri2001femtosecond,cavalleri2004evidence,wan2019optical}. Photonic applications of VO$_2$-based structures, both in the infrared~\cite{howes2020optical} and visible ranges~\cite{kepic2021optically}, have been recently developed, including the temperature control of quantum emission~\cite{szilard2019hysteresis}. 

Despite the increasing applicability of VO$_2$ in photonic devices, to the best of our knowledge its role in optical forces has never been addressed so far. With this motivation, in the present work, we investigate optical forces on oscillating electric and magnetic dipoles close to a VO$_2$ film. This system can be correlated to one composed of an excited two-level quantum emitter located at the dipole position in the dipole approximation. Therefore, the classical description alone provides newsworthy results and may predict important aspects of cavity effects on excited quantum states, a kind of approach that has been widely applied to study optical forces in such systems \cite{Novotny-book,Fortuno-2015,ChanWang-2014, Vesperinas-2010}. We consider two distinct configurations to compute optical forces: In one of them, the oscillating dipole (electric or magnetic) is parallel to the VO$_2$ film and, in the other one, it is perpendicular to the film.  Remarkably, we show that it is possible to achieve a thermal control of the optical force on the dipole and change the attractive/repulsive character of the force in the near field on both kinds of dipoles just by varying the temperature. Hence we conclude that thermal hysteresis clearly shows up in optical forces. Our results expand the applicability of phase-change materials, VO$_2$ in particular, to the external tuning of optical forces, and in general controlling light-matter interactions at the nanoscale. 

This paper is organized as follows. In the next section, we present the methodology employed to model the VO$_2$ medium and its metal-insulator phase transition and to calculate the classical expressions for the optical force near a planar surface for both configurations of dipoles. Section \ref{SecResults} comprises our main results, whereas Section \ref{SecConclusions} is dedicated to our final comments and main conclusions.


\section{Methodology}
\label{SecMethods}

\begin{figure}[H]

	\centering
	\includegraphics[width=0.5\linewidth]{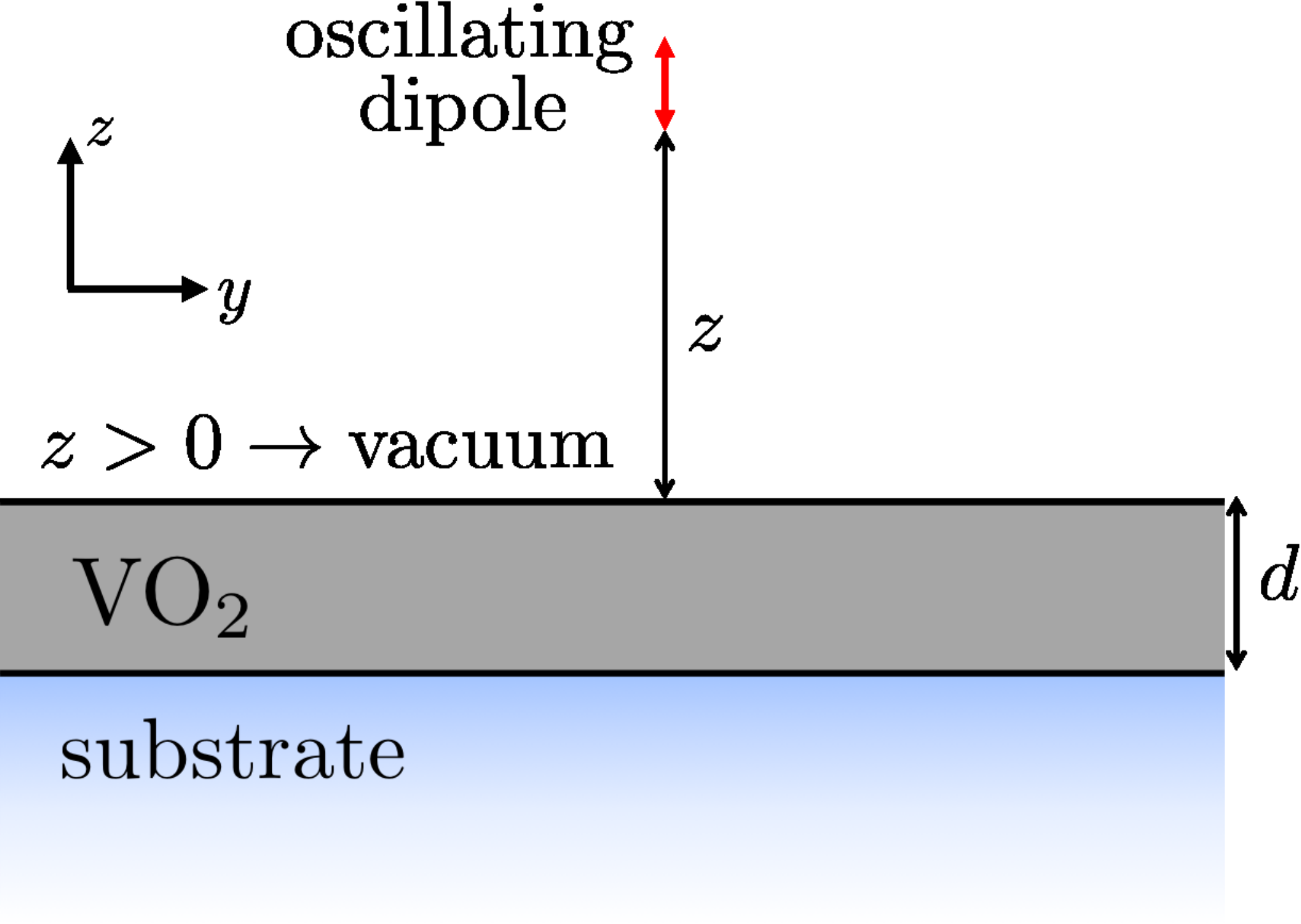}
	\caption{Schematic representation of the system. An oscillating dipole in vacuum is placed at a distance $z$ of a VO$_2$ film of thickness $d$ suported by a Al$_2$O$_3$ substrate.
	\label{Fig:systemVO2}}

\end{figure}

In order to study optical forces, we shall consider a prescribed oscillating dipole near a VO$_2$ film of thickness $d = 200$ nm, supported by a saphire substrate (Al$_2$O$_3$). The oscillating dipole can be either electric or magnetic and it is placed at a distance $z$ from the surface as shown in Figure~\ref{Fig:systemVO2}.

In the following subsections we introduce the theoretical approach employed to characterize the film and its metal-insulator phase transition. Next, we describe the main equations used to evaluate the optical forces on the dipole.


\subsection{Modelling the VO$_2$ metal-insulator transition}


In this work, we used the approach developed in Refs.~\cite{kort2018passive,szilard2019hysteresis} to model VO$_2$ optical properties in terms of its dielectric function, applying the Bruggeman effective medium theory (BEMT) \cite{Bruggeman-1935}. In short, VO$_2$ presents an insulating behavior and monoclic structure \cite{Imada-1998} up to a critical temperature ${T_{\rm MIT} \sim 340}$~K at which it undergoes a metal-insulator transition that consists of a structural phase transition to a rutile-type structure. In practice, however, the transition is smooth and at temperatures close to $T_{\rm MIT}$, the material presents a mixture of both metallic and insulating domains \cite{Qazilbash-2007, Qazilbash-2009}, which can be modeled as spheroidal metalic inclusions in a dielectric host medium. It is therefore possible to treat VO$_2$ as a two-phase system with an effective permittivity depending on the permittivities of both metallic and insulating regions and their respective volume filling fractions $f$ and $1-f$ ($0< f < 1$) \cite{Choy-book}. Here, the filling fractions are modeled as a logistic function of the temperature to emulate the histeretical behavior of the VO$_2$ electric response as the temperature is ramped up or down \cite{szilard2019hysteresis}.

The task now is to calculate the VO$_2$ effective dielectric constant $\varepsilon_{\text{VO}_2} (\lambda, T)$ for temperatures in which the heteregeneous regime is present, i. e., close to $T_{\rm MIT}$. In the framework of BEMT, this may be be obtained from \cite{szilard2019hysteresis}
\begin{align}
	(1-&f) \left\{\frac{\varepsilon_{\rm d} - \varepsilon_{\text{VO}_2}}{\varepsilon_{\text{VO}_2} + L \,(\varepsilon_{\rm d}-\varepsilon_{\text{VO}_2})} + \frac{4(\varepsilon_{\rm d} - \varepsilon_{\text{VO}_2})}{2\,\varepsilon_{\text{VO}_2} + (1-L)(\varepsilon_{\rm d}-\varepsilon_{\text{VO}_2})}\right\} \nonumber \\
	+&f\left\{\frac{\varepsilon_{\rm m} - \varepsilon_{\text{VO}_2}}{\varepsilon_{\text{VO}_2}+ L \, (\varepsilon_{\rm m}-\varepsilon_{\text{VO}_2})} + \frac{4(\varepsilon_{\rm m} - \varepsilon_{\text{VO}_2})}{2\,\varepsilon_{\text{VO}_2} + (1-L)(\varepsilon_{\rm m}-\varepsilon_{\text{VO}_2})}\right\} =0 \,,
\label{eq:BEMTVO2} 
\end{align}
where $L$ ($0\leq L \leq 1$) is the depolarization factor related to the shape of the metallic inclusions, and wavelength and temperature dependences have been omitted for simplicity. Also, $\varepsilon_{\rm d}$ and $\varepsilon_{\rm m}$ denote the dielectric constants of VO$_2$ in the purely insulating ($T\ll T_{\rm MIT}$, $f = 0$) and in the purely metallic ($T\gg T_{\rm MIT}$, $f = 1$) phases, respectively. These quantities are given by
\begin{eqnarray}
	\varepsilon_{\rm m} (\omega) &=& 1 - \frac{\omega_{{\rm m}}^2}{\omega^2 + i \, 
\gamma_{\rm m} \, \omega} \,, 
\label{eq:DrudeMet}\\
	\varepsilon_{\rm d} (\omega) &=& 1 + \frac{\omega_{\rm d}^2}{\omega_{\rm R}^2 \,-\,\omega^2 - i \, \omega \, \gamma_{ \, \rm d}} \,,
\label{eq:DrudeDie}
\end{eqnarray}
where $\omega_{\rm R}$ is the material resonance frequency, $\omega_{\rm m} \, ({\omega_{\rm d}})$ is the the plasma frequency (oscillating strength), and $\gamma_{\rm m} \,(\gamma_{\rm d})$ is the inverse of the relaxation time of the metallic phase (insulator phase). These parameters were obtained by numerically fitting the experimental data reported in \cite{peterseim2016optical} in the range $1$ $\mu$m $\leq \lambda \leq 10$ $\mu$m.


\subsection{Optical forces near a planar surface}


In this subsection we shall briefly establish the expressions for the optical forces acting on a harmonically oscillating electric or magnetic dipole, when they are near a given planar surface whose reflection Fresnel coefficients for the transverse electric (TE) and transverse magnetic (TM) waves are known. Hence, the electromagnetic fields responsible for the optical force acting on the dipole will be the scattered part of the own dipole field. The optical force acting on a time-dependent electric dipole ${\bm d}(t)$ located at a generic position ${\bm r}_0$ is given by \cite{gordon1973radiation,chaumet2000}
\begin{equation}
{\bm F}^e(t) = ({\bm d}(t)\cdot\nabla){\bm E}({\bm r}_0,t)  + \frac{1}{c}\dot{\bm d}(t)\times{\bm B}({\bm r}_0,t) \,.
\end{equation}
For the case at hand, the dipole and the electromagnetic fields have the same harmonic time-dependence, namely,
\begin{equation}
 {\bm d}(t) = {\bm d}_0 e^{-i\omega t}\; ;\;\;\;\;  {\bm E}({\bm r},t) = {\bm E}_0({\bm r}) e^{-i\omega t}\; ;\;\;\;\; 
 \mbox{and}\;\;\;\; {\bm B}({\bm r},t) = {\bm B}_0({\bm r}) e^{-i\omega t}\, .
\end{equation}
The time average of the force is given by
\begin{eqnarray}
\langle{\bm F}^e\rangle &=& \langle (\mathcal{R}e\,{\bm d}\cdot\nabla) \mathcal{R}e\,{\bm E}\rangle + \frac{1}{c} \langle \mathcal{R}e\,{\dot{\bm d}}\times\, \mathcal{R}e\,{\bm B}\rangle\cr\cr
&=& 
\frac{1}{2}\mathcal{R}e \Bigl\{({\bm d}_0\cdot\nabla){\bm E}_0^*  - \frac{i\omega}{c} {\bm d}_0\times{\bm B}_0^*   \Bigr\} \,.
\end{eqnarray}
For the $i$-component of the force, we have
\begin{equation}
\langle{F^e_i}\rangle = \frac{1}{2}\mathcal{R}e \Bigl\{d_{0j}\partial_j E_{0i}^*  - \frac{i\omega}{c}\epsilon_{ijk}d_{0j}B_{0k}^*   \Bigr\} \,.
\end{equation}
Throughout this paper, we adopted the implict sum notation for repeted indices. Using Faraday's law, ${\bm B}_0 = \frac{c}{i\omega} \nabla\times{\bm E}_0$, so that $B_{ok}^* = \frac{ic}{\omega}\epsilon_{l n k}\partial_l E_{0n}^*$, as well as the identity $\epsilon_{ijk}\epsilon_{l n k} = \delta_{il}\delta_{j n} - \delta_{in}\delta_{jl}$, we finally obtain \cite{chaumet2000}
\begin{equation}
{F^e_i}({\bm r}_0) = \frac{1}{2}\mathcal{R}e \Bigl\{d_{0j}\partial_i  E_{0j}^*({\bm r}_0)\Bigr\}\, ,
\end{equation}
where we brought back the dipole position ${\bm r}_0$ and, for convenience of notation, we wrote the time average $\langle{F^e_i}\rangle ({\bm r}_0)$ simply as $F^e_i({\bm r}_0)$.

Analogously, it can be shown that the time average force acting on an oscillating magnetic dipole ${\bm m}(t)  = \mathcal{R}e\,\left( {\bm m}_0 e^{-i \omega t}\right)$ is given by
 $F^m_i ({\bm r}_0)= \frac{1}{2} \, \mathcal{R}e \, \bigl\{   \mu_0 \, m_{0j} \, \partial_i \, H^{0*}_j ({\bm r}_0) \bigr\}$, so that if the two dipole moments are present, the total force on the particle is
\begin{equation}
{F^e_i}({\bm r}_0)  + F^m_i ({\bm r}_0) = \frac{1}{2} \, \mathcal{R}e \, \Bigl\{ d_{0j}\partial_i  E_{0j}^*({\bm r}_0) + \mu_0 \, m_{0j} \, \partial_i \, H^{0*}_j ({\bm r}_0) \Bigr\} \,.
\label{eq:OptForceVO2}
\end{equation}

Now, let us consider that the dipole is near a planar surface. In this case, in order to compute the optical force on the dipole, we need to take into account in the previous equations the electromagnetic field that is scattered by the neighboring surface and acts back in the dipole. As usual, these scattered electromagnetic fields can be calculated with the aid of the corresponding Green function $\mathbb{G} ({\bm r}, {\bm r}^\prime; \omega)$.

Let us first compute the electromagnetic force on an emitter at position ${\bm r}_0$ with only electric dipole transition. Adopting ${\bm m} = 0$ in the previous equation the time average of the electric force reads
\begin{equation}
{\bm F}^e ({\bm r}_0) = \frac{1}{2} \, \mathcal{R}e \, \Bigl\{ d_{j} \, \nabla E_{j}^{*} ({\bm r}_0) \Bigr\} \,,
\label{ForcaEl}
\end{equation}
where, for simplicity of notation, we are omitting the subscripts $0$ in the electric dipole moment as well as in the electric field. In order to compute the force on the electric dipole using the previous equation, we need the scattered electric field, which can be written in terms of the scattered Green function as ${\bm E}^{(S)} ({\bm r}) = \omega^2 \mu_0 \, \mathbb{G}_e^{(S)} ({\bm r}, {\bm r}_0; \omega) \cdot {\bm d}$. The scattered Green function for a planar geometry is well-known in the literature \cite{Novotny-book}, so that last equation leads to
\begin{align}
F^e_l &= \frac{1}{2} \, \omega^2 \, \mu_0 \,  \mathcal{R}e \Bigl\{ d_i^* \, d_j \,  \partial_l\,  G_{ij}^{(S)} (z,z_0; \omega)\Bigr\}_{z=z_0} \nonumber \\
&= - \frac{1}{4} \, \omega^2 \, \mu_0 \,  \mathcal{R}e \left\{ d_i^* d_j \, \int_{0}^{\infty} dk_\parallel \, e^{2i \, k_{z0} z_0} \int_0^{2 \pi}  \frac{d \varphi}{(2 \pi)^2}  \, \frac{k_\parallel k_l}{k_{z0}} \, R_{ij} \right\}
\end{align}
for the $l$-component of the force, with $k_{z0} = \sqrt{k_0^2-k_\parallel^2}$, $k_0 = \omega/c$ and $\mathbb{R}$ given by
\begin{equation}
	\mathbb{R} = \sum_{\rm p, q = \{ TE, TM \}} r^{\rm p, q} \, {\bm \epsilon}_{\rm p}^{+} \otimes {\bm \epsilon}_{\rm q}^{-} \,,
\end{equation}
where $r^{\rm p, q}$ (${\rm p, q} =$ TE, TM) are the usual Fresnel reflection coefficients for a p-polarized incident wave being scattered into a q-polarized reflected wave, and ${\bm \epsilon}_{\rm TE}^{\pm}$ and ${\bm \epsilon}_{\rm TM}^{\pm}$ denote the TE- and TM-polarization unitary vectors, respectively \cite{Novotny-book}. It can be shown that
\begin{align}
F^e_{\rm x} (z) & = \frac{1}{8 \pi \varepsilon_0} \mathcal{R}e \int_{0}^{\infty} d k_\parallel \,   \Bigl[ \; i \, r^{\rm TM, TM}  \, \mathcal{I}m (d_{\rm x}^* \, d_{\rm z})  + \frac{k_0}{k_{z0}} \, r^{\rm TE, TM} \, \mathcal{R}e \, (d_{\rm y}^* \, d_{\rm z})  \Bigr] k_\parallel^3 \, e^{2i \, k_{z0} z}  \,, \\ 
F^e_{\rm z} (z) &= \begin{aligned}[t] 
\label{eq:FxFzElectric}
  -\frac{1}{16 \pi \varepsilon_0} \, \mathcal{R}e \int_{0}^{\infty} d k_\parallel \, \Bigl[ &|d_\parallel|^2 \left( k_0^2 \, r^{\rm TE, TE} - k_{z0}^2 \, r^{\rm TM, TM} \right) + 2\,|d_{\rm z}|^2 \, k_\parallel^2 \, r^{\rm TM, TM} \\
&+ 4 \, i \; \mathcal{I}m (d_{\rm x}^* \, d_{\rm y}) \, k_0\, k_{z0}\, r^{\rm TE, TM} \Bigr] k_\parallel  \, e^{2i \, k_{z0} z} \,.
\end{aligned}
\end{align}
In the previous equations, we have already performed the angular integral in $d \phi$ and $|d_\parallel|^2 = |d_{\rm x}|^2 + |d_{\rm y}|^2$; $F^e_{\rm y}$ is the analogous to $F^e_{\rm x}$ provided we replace the dipole moment components properly; meaning, $d_{\rm x} \rightarrow d_{\rm y}$ and $d_{\rm y} \rightarrow - d_{\rm x}$.

For an isotropic material like VO$_2$, $r^{\rm TE, TM} = 0$. In this work, we will only be concerned with cases in which the dipole components do not present a relative phase between them and can be considered real. Consequently, $\mathcal{I}m (d_{\rm x}^* d_{\rm z}) = \mathcal{I}m (d_{\rm x}^* d_{\rm y}) = \mathcal{I}m (d_{\rm y}^* d_{\rm z}) = 0$ and the optical forces acting on the dipoles will have only a component perpendicular to the VO$_2$ film.

It is convenient to split the vertical force into two contributions, namely: $F^e_{\perp}$, which is proportional to the component of the electric dipole moment perpendicular to the surface $d_{\rm z}$, and $F^e_{\parallel}$, which is proportional to the component of the electric dipole moment parallel to the surface $d_{\parallel}$. These contributions can be written as
\begin{align}
	\frac{F^e_{\perp}}{\Gamma_0^e \,  \hbar k_0 } \, (z)&= - \frac{3}{8 k_0^4} \, \frac{|d_z|^2}{|{\bm d}|^2} \, \mathcal{R}e \int_0^{\infty} d k_\parallel \, r^{\rm TM, TM} \,  k_\parallel^3 \, e^{2i \, k_{z0} z} \,,
\label{eq:FzEPerp}  \\
	\frac{F^e_{\parallel}}{\Gamma_0^e \,  \hbar k_0 } \, (z) &= - \frac{3}{16 k_0^4} \, \frac{|d_\parallel|^2}{|{\bm d}|^2} \, \mathcal{R}e \int_0^{\infty} d k_\parallel \, \left( - k_{z0}^2 \, r^{\rm TM, TM} + k_0^2 \, r^{\rm TE, TE} \right) \, k_\parallel \, e^{2i \, k_{z0} z}  \,.
\label{eq:FzEPar}
\end{align} 
In order to deal only with dimensionless quantities, we normalized the force by the quantity $\Gamma_0^e \, \hbar k_0$, where $\Gamma_0^e = |{\bm d}_0|^2 \,k_0^3/(3 \pi \varepsilon_0 \hbar)$ is the spontaneous emission rate of a two-level system in the empty space with transition dipole moment equal to the dipole moment ${\bm d}_0$ (the quantity $\Gamma_0^e \, \hbar k_0$ can be interpreted as the recoil force on a quantum emitter whose transition frequency is $\omega_0$). Recall that, if we take the average in all possible orientations, for an isotropic emitter $|d_z|^2/|{\bm d}|^2 = 1/3$ and $|d_\parallel|^2/|{\bm d}|^2 = 2/3$.

Let us now turn our attention to the calculation of the optical force on a magnetic oscillating dipole. The procedure to compute this force follows the same steps as those for the electric dipole case. The main difference is that we need now the magnetic Green function $\mathbb{G}_m ({\bm r}, {\bm r}_0; \omega)$, instead of the electric one. With this is mind,  the optical force on the magnetic dipole reads
\begin{equation}
{\bm F}^m ({\bm r}_0) = \frac{\mu_0}{2} \, \mathcal{R}e \, \Bigl\{ m_{j} \, \nabla H_{j}^{*} ({\bm r}_0) \Bigr\} \,.
\end{equation}
As in Equation (\ref{ForcaEl}), we are omitting the subscripts $0$ in the magnetic dipole moment, as well as in the magnetic field. The scattered part of the magnetic field at a generic position ${\bm r}$ created by a magnetic dipole which is located at position ${\bm r}_0$ is given by ${\bm H}^{(S)} ({\bm r}) = \mathbb{G}_{m}^{(S)} ({\bm r}, {\bm r}_0; \omega) \cdot {\bm m}$.
The magnetic Green function can be written as $\mathbb{G}_m^{(S)} ({\bm r}, {\bm r}'; \omega) = \mu_0^{-1} \, \overrightarrow{\nabla} \times \mathbb{G} ({\bm r}, {\bm r}'; \omega) \times \overleftarrow{\nabla}'$~\cite{Buhmann-bookII}. For a planar geometry, we find the expressions for the optical force on a magnetic oscillating dipole in terms of the Fresnel reflection coefficients, namely
\begin{align}
F^m_{\rm x} (z) & = \frac{\mu_0}{8 \pi} \mathcal{R}e \int_{0}^{\infty} d k_\parallel \, \Bigl[ \; i \, r^{\rm TE, TE}  \, \mathcal{I}m (m_{\rm x}^* \, m_{\rm z})  - \frac{k_0}{k_{z0}} \, r^{\rm TE, TM} \, \mathcal{R}e \, (m_{\rm y}^* \, m_{\rm z})  \Bigr] \, k_\parallel^3 \, e^{2i \, k_{z0} z} \,,  \\ 
F^m_{\rm z} (z) &= \begin{aligned}[t] 
\label{eq:FxFzMagnetic}
  -\frac{\mu_0}{16 \pi} \, \mathcal{R}e \int_{0}^{\infty} d k_\parallel \, \Bigl[ &|m_\parallel|^2 \left(k_0^2 \, r^{\rm TM, TM} - k_{z0}^2 \, r^{\rm TE, TE} \right) + 2\,|m_{\rm z}|^2 \, k_\parallel^2 \, r^{\rm TE, TE} \\
&- 4 \, i \; \mathcal{I}m (m_{\rm x}^* \, m_{\rm y}) \, k_0\, k_{z0}\, r^{\rm TE, TM} \Bigr]\,  k_\parallel \, e^{2i \, k_{z0} z} \,.
\end{aligned}
\end{align}
Comparing the above formulas with those written in Equation (\ref{eq:FxFzElectric}), we see that they can be mapped one into another when we make the replacements $d_i \rightarrow m_i$, $r^{\rm TM, TM} \rightarrow r^{\rm TE, TE} $, $r^{\rm TE, TE} \rightarrow r^{\rm TM, TM} $ and $r^{\rm TE, TM} \rightarrow - r^{\rm TE, TM} $.
 
Finally, as we have done for the optical force on the oscillating electric dipole, it is also convenient  to split the vertical force into two contributions, to wit: $F^m_{\perp}$ (proportional to the component of the electric dipole moment perpendicular to the surface $m_{\rm z}$) and $F^m_{\parallel}$ (proportional to the component of the electric dipole moment parallel to the surface $m_{\parallel}$). For an isotropic medium, they are given by
\begin{align}
	\frac{F^m_{\perp}}{\Gamma_0^m \,  \hbar k_0 } \, (z)&= - \frac{3}{8 k_0^4} \, \frac{|m_z|^2}{|{\bm m}|^2} \, \mathcal{R}e \int_0^{\infty} d k_\parallel \, r^{\rm TE, TE} \,  k_\parallel^3 \, e^{2i \, k_{z0} z} \,,
\label{eq:FzMPerp} \\
	\frac{F^m_{\parallel}}{\Gamma_0^m \,  \hbar k_0 } \, (z) &= - \frac{3}{16 k_0^4} \, \frac{|m_\parallel|^2}{|{\bm m}|^2} \, \mathcal{R}e \int_0^{\infty} d k_\parallel \, \left(- k_{z0}^2 \, r^{\rm TE, TE} + k_0^2 \, r^{\rm TM, TM} \right) \,  k_\parallel \, e^{2i \, k_{z0} z} \,,
\label{eq:FzMPar}
\end{align} 
where we applied an analogous normalization to the one used in Equations (\ref{eq:FzEPerp}) and (\ref{eq:FzEPar}), but now with $\Gamma^m_0 = \mu_0 \, k_0^3  \, |{\bm m}|^2/(3 \pi \, \hbar )$ being the spontaneous emission rate of a magnetic emitter in empty space.
\end{paracol}

\begin{paracol}{2}
\switchcolumn


\section{Results and discussions}
\label{SecResults}


We now proceed to detailed discussions of our results regarding the optical forces acting on oscillating dipoles close to a VO$_2$ film. As previously mentioned, we performed our analysis for the cases of electric and magnetic dipoles and each of the following subsections accounts for one of them.


\subsection{Electric Dipole}


In Figure~\ref{fig:FT_ED}, we plot the force on the oscillating electric dipole in the perpendicular configuration, due to the presence of VO$_2$ medium near the MIT, as a function of temperature. We chose $z = 50$ nm, so that we are in the near-field regime, and the electric dipole is oscillating perpendicularly to the film. As expected, the material thermal hysteresis is directly reflected in these curves. The most compelling feature that can be noted is the fact that there are some values of wavelengths $\lambda$ for which the attractive/repulsive character of the force may be interchanged just by heating or cooling the VO$_2$ film.

\begin{figure}[h!]

	\centering
	\includegraphics[width=0.95\linewidth]{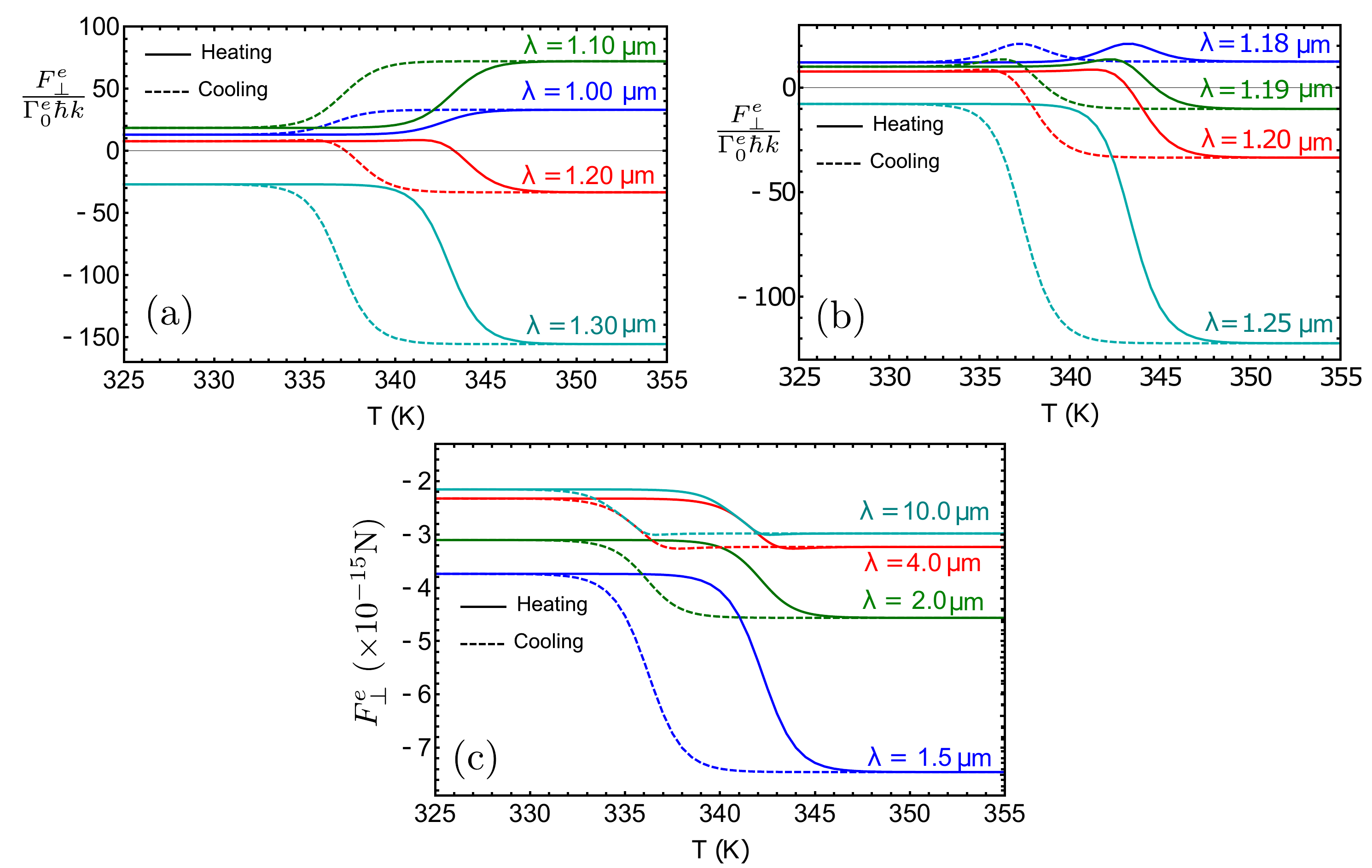}	
	\caption{Panels {\bf (a)} and {\bf (b)}: Force on the electric dipole in the perpendicular configuration normalized by $\Gamma^e_0\hbar k$ as a function of temperature $T$ for different values of wavelengths $\lambda$. The dipole is assumed to be at a distance $z = 50$~nm from the VO$_2$ film. In panel {\bf (c)}, we considered a dipole moment $d_0 \sim 3\times 10^{-27}$ Cm.}
	\label{fig:FT_ED}

\end{figure}

In order to furtherly exploit this property, Figure~\ref{fig:FLambda_ED} displays the force dependence with the wavelength $\lambda$ of the radiation emitted by a dipole at the same distance $z~=~50$~nm. Note that, due to the thermal hysteresis, two distinct temperatures can correspond to the same curve in Figure~\ref{fig:FLambda_ED} depending on whether the system is being heated or cooled. It can be spotted that the change in the force sign with temperature occurs only for the short range $1.186$ $\mu$m $< \lambda < 1.228$ $\mu$m (where the blue and the purple lines cross $F_\perp^e = 0$). For higher wavelengths, the force is always attractive, whereas, for smaller wavelengths, always repulsive. It may also be noted from Figure~\ref{fig:FLambda_ED}(a) that, as the wavelength increases, the force does not significantly change with $\lambda$ for a fixed temperature. This behavior can also be inferred from Figure~\ref{fig:FT_ED}(c), which reveals that the relative change in $F_\perp^e$ at a given phase reduces with $\lambda$. This remark should be considered carefully. We normalized Equations (\ref{eq:FzEPerp}) and (\ref{eq:FzEPar}) by $\Gamma_0^e \, \hbar k$ to keep our results as general as possible, so that they do not explicitly depend on any particular value of the dipole moment $|{\bm d}|$ (just on the ratio $d_z/|{\bm d}|$). However, in doing so, $\Gamma_0^e \, \hbar k \sim 1/\lambda^4$ and our normalized results artificially grow with $\lambda^4$. To avoid any misleading conclusions, in Figures~\ref{fig:FT_ED}(c) and \ref{fig:FLambda_ED}(a), we plot the non-normalized $F_\perp^e$ considering large $\lambda$ intervals so that this artificial  $\lambda$-dependence is absent. In addition, we assume a dipole moment $d_0 \sim 3 \times 10^{-27}$ Cm. This corresponds to a transition wavelength of $1$ $\mu$m and a static polarizability of $4\pi\epsilon_0 r^3$, with $r = 10$ nm.

\begin{figure}[h]

	\centering
	\includegraphics[width=0.95\linewidth]{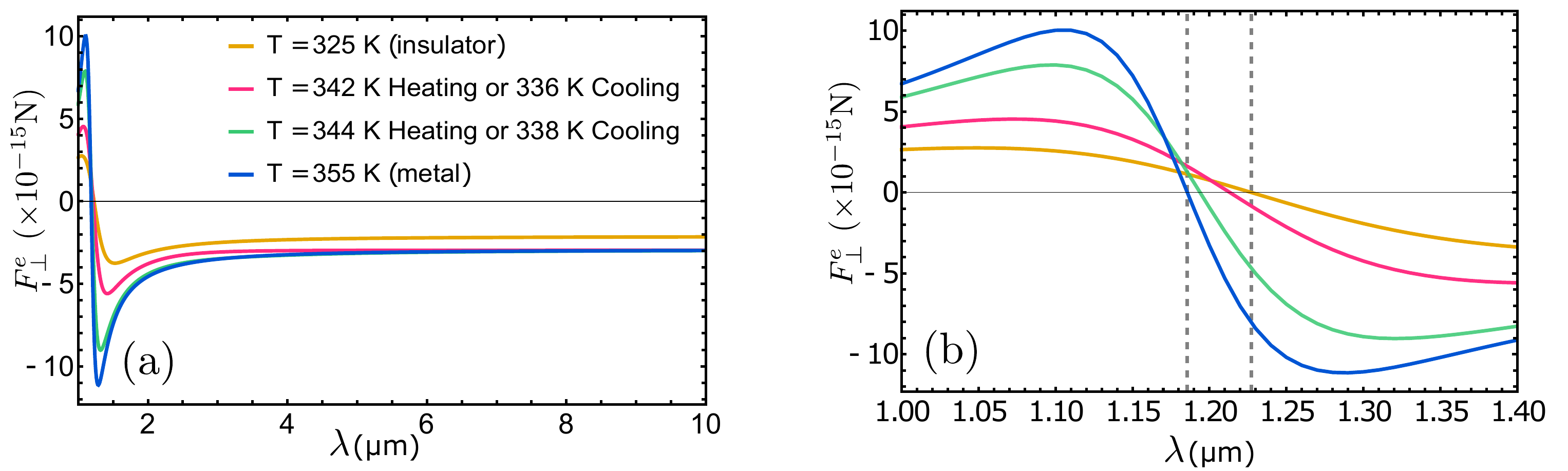}
	\caption{{\bf (a)} Force on the electric dipole (with dipole moment $d_0 \sim 3\times 10^{-27}$ Cm) in the perpendicular configuration as a function of wavelength $\lambda$ for different temperatures $T$. The dipole is assumed to be at a distance $z = 50$ nm from the VO$_2$ film. {\bf (b)} Details of the previous panel in the range $1$ $\mu$m $< \lambda < 1.4$ $\mu$m. Dashed lines indicate the $\lambda$-interval for which the force may change its sign depending on the temperature.}
	\label{fig:FLambda_ED}

\end{figure}

Our last remarks can be enlightened in terms of the near-field regime of Equation $\;$ (\ref{eq:FzEPerp}). The quasi-static limit, obtained by taking $c \rightarrow \infty$, leads us to
\begin{align}
	F_\perp^e  \simeq - \frac{1}{16 \pi \varepsilon_0} \, \frac{3 \, d_{z, \parallel}^2}{4 z^4} \,\, \eta^e, && \eta^e = \frac{|\varepsilon_{\textrm{VO}_2}|^2 - 1}{|\varepsilon_{\textrm{VO}_2} + 1|^2} \,.
\label{eq:ForceNF_ED}
\end{align}
Figure~\ref{fig:ForcefactorED} shows the dependence on the factor $\eta^e$ with $\lambda$ for different temperatures. By comparing it with Figure~\ref{fig:FLambda_ED}, one realizes that it exhibits exactly the same behavior, except for the negative sign. In particular, the force changes sign at wavelengths for which $|\varepsilon_{\textrm{VO}_2}| \sim 1$, i. e., $\lambda \sim 1.2$ $\mu$m. Around this value, there is a peak on the factor $\eta^e$, responsible for the non-monotonic behavior with $\lambda$ appearing in Figure~\ref{fig:FT_ED}(a). Note that, for wavelengths $\lambda \gtrsim 2$ $\mu$m, this factor is basically independent of $\lambda$ and its value in the dielectric and metallic phases differs only by a fixed value. These features can be explained due to the behavior of Drude and Drude-Lorenz permittivities $\varepsilon_{\rm m}$ and $\varepsilon_{\rm d}$ [Equations~(\ref{eq:DrudeMet}) and (\ref{eq:DrudeDie})] and it is also present in Figure~\ref{fig:FLambda_ED}(b). The permittivities approach a fixed value as $\omega \rightarrow 0$. Similar conclusions occur for the case in which the electric dipole oscillates parallel to the film. In fact, the near-field regime of Equation (\ref{eq:FzEPar}) furnishes $F_\perp^e \simeq 2\, F_\parallel^e$, so that $F_\parallel^e$ only differs from the perpendicular contribution by a factor of two. This factor can be physically interpreted in terms of the field created by a static electric dipole. At a given distance from the dipole, this field has its modulus along the dipole direction twice as large as its modulus at the same distance along a direction perpendicular to dipole.

\begin{figure}[h]

	\centering
	\includegraphics[width=0.95\linewidth]{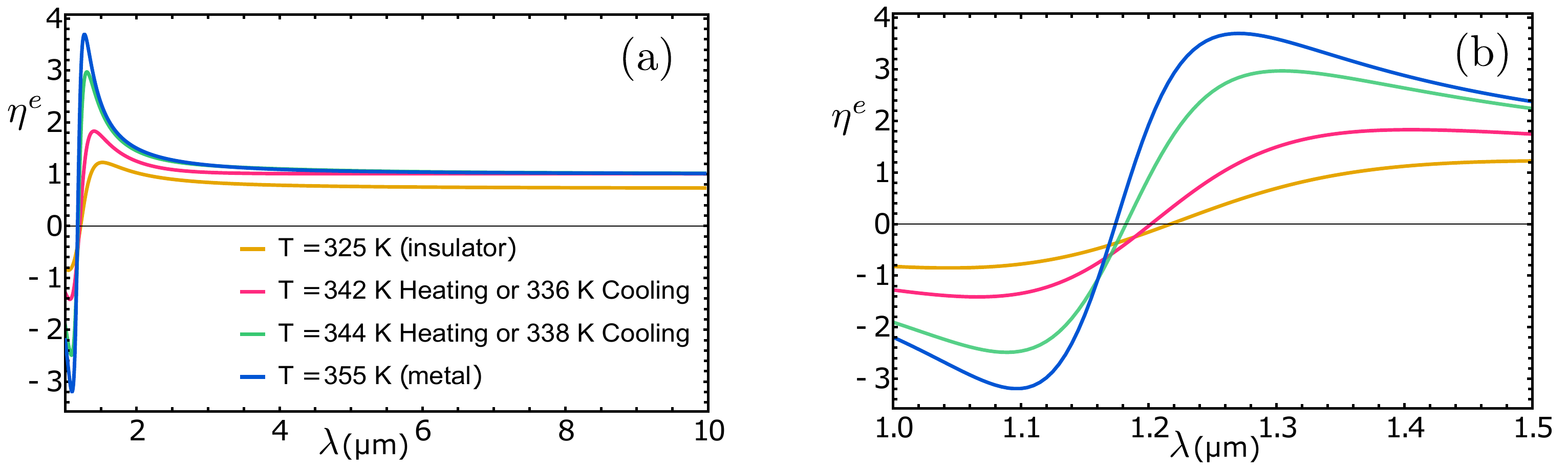}
	\caption{{\bf (a)} Factor $\eta^e$ as a function of wavelengths for different temperatures. {\bf (b)} Details of the previous panel in the range $1$ $\mu$m $< \lambda < 1.5$ $\mu$m.}
	\label{fig:ForcefactorED}

\end{figure}

Figure~\ref{fig:Fz325K355K_ED} displays the normalized force $F_\perp^e/\Gamma_0^e \hbar k$ as a function of $z$ at the insulator and metal phases for different wavelengths $\lambda$. Note that the force oscillates around zero for large distances and diverges with $\sim 1/z^4$ in the near-field regime. Curiously, for certain wavelengths $\lambda \sim 1.2$ $\mu$m, the force may diverge positively or negatively depending on the phase in which the VO$_2$ film is found. For instance, in the insulator phase [Figure~\ref{fig:Fz325K355K_ED}(a)], $F_\perp^e$ diverges positively for $\lambda = 1.2$ $\mu$m, while, in the metallic phase [Figure~\ref{fig:Fz325K355K_ED}(b)], it diverges negatively. Conversely, the force $F_\perp^e$ for the others $\lambda$ shown in Figure~\ref{fig:Fz325K355K_ED} do not change qualitatively with the temperature.

\begin{figure}[h]

	\centering
	\includegraphics[width=0.95\linewidth]{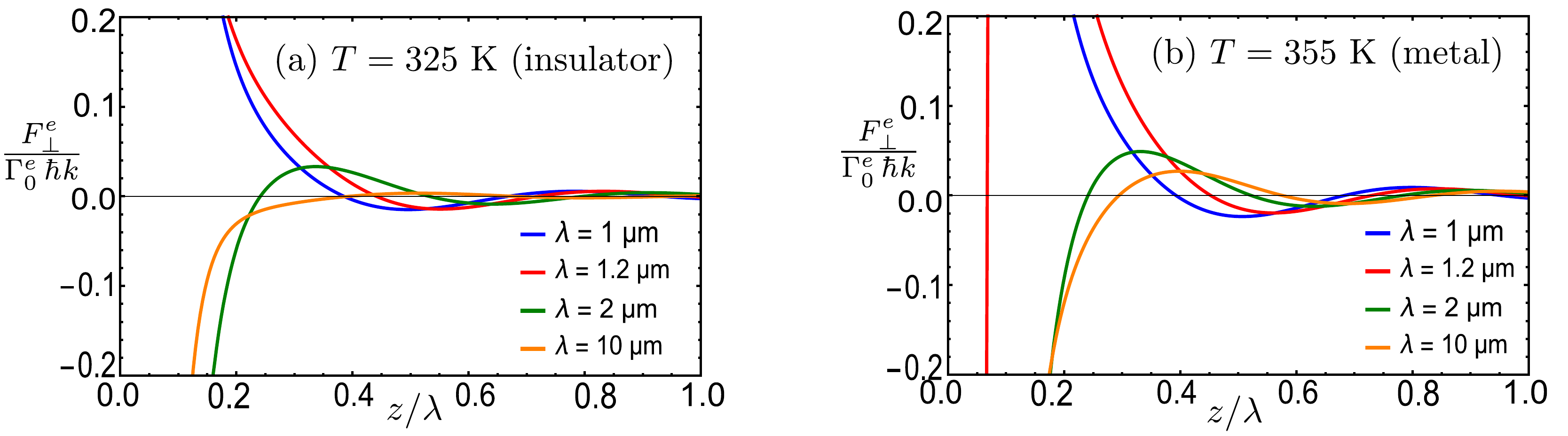}
	\caption{Force $F_\perp^e$ in the perpendicular configuration normalized by $\Gamma^e_0\hbar k$ as a function of the distance at the insulator [panel {\bf (a)}] and metallic phases [panel {\bf (b)}] for different wavelengths.}
	\label{fig:Fz325K355K_ED}

\end{figure}

In order to unveil such behavior, Figure~\ref{fig:FzLambdaT_ED} illustrates how the force $F_\perp^e$ varies as a function of the distance $z$ for different temperatures at fixed wavelengths $\lambda \sim 1.2$ $\mu$m. For example, considering $z \lesssim 50$ nm in pannels (b) and (c), $F_\perp^e$ can be either attractive or repulsive depending on the temperature, but the same behavior is not verified in the other pannels. One may notice that, as $\lambda$ increases, the force at very small distances changes gradually from repulsive to attractive.  Hence, the force may diverge negatively or positively as $z \rightarrow 0$ depending on $\lambda$ and also on the temperature. Actually, this type of behavior was already expected from Figure~\ref{fig:FT_ED}(b), when we verified that at a fixed distance $z = 50$ nm the force could change sign just through heating for wavelenghts around $\lambda \sim 1.2$ $\mu$m. Nevertheless, in studying this change of behavior as a function of $z$, one may find results that are robust with respect to the distance $z$. In particular, Figure~\ref{fig:FzLambdaT_ED} reveals a new property, to wit, the fact that the first equilibrium point of the system changes its stability character with the temperature.

\begin{figure}[h]

	\centering
	\includegraphics[width=0.95\linewidth]{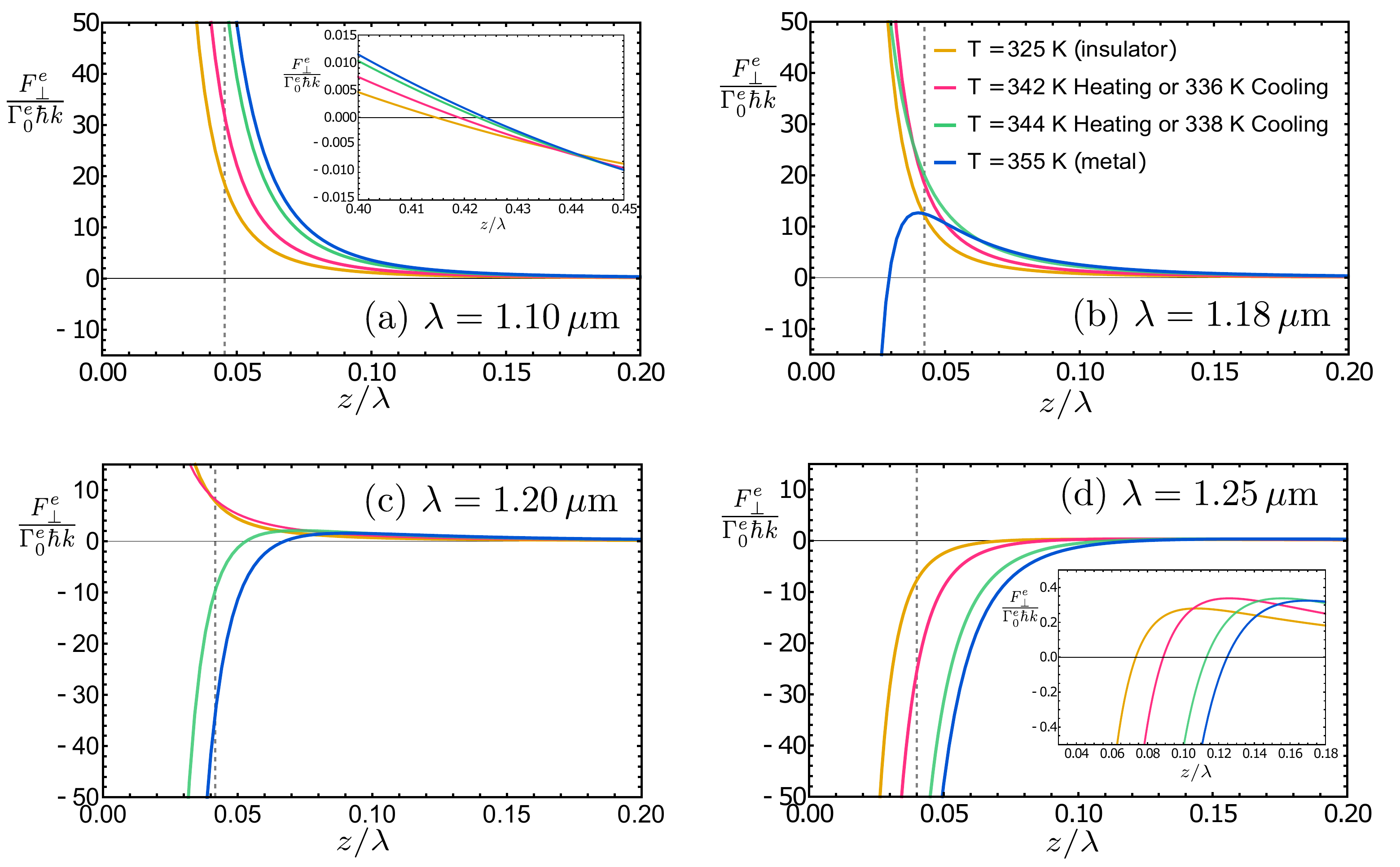}
	\caption{Force $F_\perp^e$ in the perpendicular configuration as a function of the distance for different temperatures and fixed wavelengths $\lambda = 1.10$ $\mu$m [panel {\bf (a)}], $\lambda = 1.18$ $\mu$m [panel {\bf (b)}], $\lambda = 1.20$~$\mu$m [panel {\bf (c)}] and $\lambda = 1.25$ $\mu$m [panel {\bf (d)}]. The insets in panels {\bf (a)} and {\bf (d)} show the intercept of these plots with $x$-axis.}
	\label{fig:FzLambdaT_ED}

\end{figure}

To investigate further this feature, Figure~\ref{fig:StatPoint_ED} shows the normalized position $z_{\rm eq}/\lambda$ of the first equilibrium point as a function of $\lambda$ at fixed temperatures. Note that, for $1.17$ $\mu$m $< \lambda < 1.21$ $\mu$m, $z_{\rm eq}/\lambda$ varies discontinuously with $\lambda$, and the position of the discontinuity depends on the temperature. In fact, this discontinuous behavior is a consequence of the fact that the equilibrium point changes from unstable to stable as $\lambda$ increases. Moreover, for any $\lambda$ in between the aforementioned range, it is always possible to place the dipole at a distance $z$ at which the force will change sign just by heating (or cooling) the VO$_2$ film. For instance, the force $F_\perp^e$ on the dipole is repulsive at a distance $z = 50$ nm for $\lambda = 1.18$ $\mu$m regardless the temperature [see Figures~\ref{fig:FT_ED}(b) and \ref{fig:FzLambdaT_ED}(b)], but it changes sign for smaller $z$. In contrast, the force is always attractive for  $\lambda = 1.25$ $\mu$m regardless the distance $z$ and the temperature of the film.

\begin{figure}[h]

	\centering
	\includegraphics[width=0.85\linewidth]{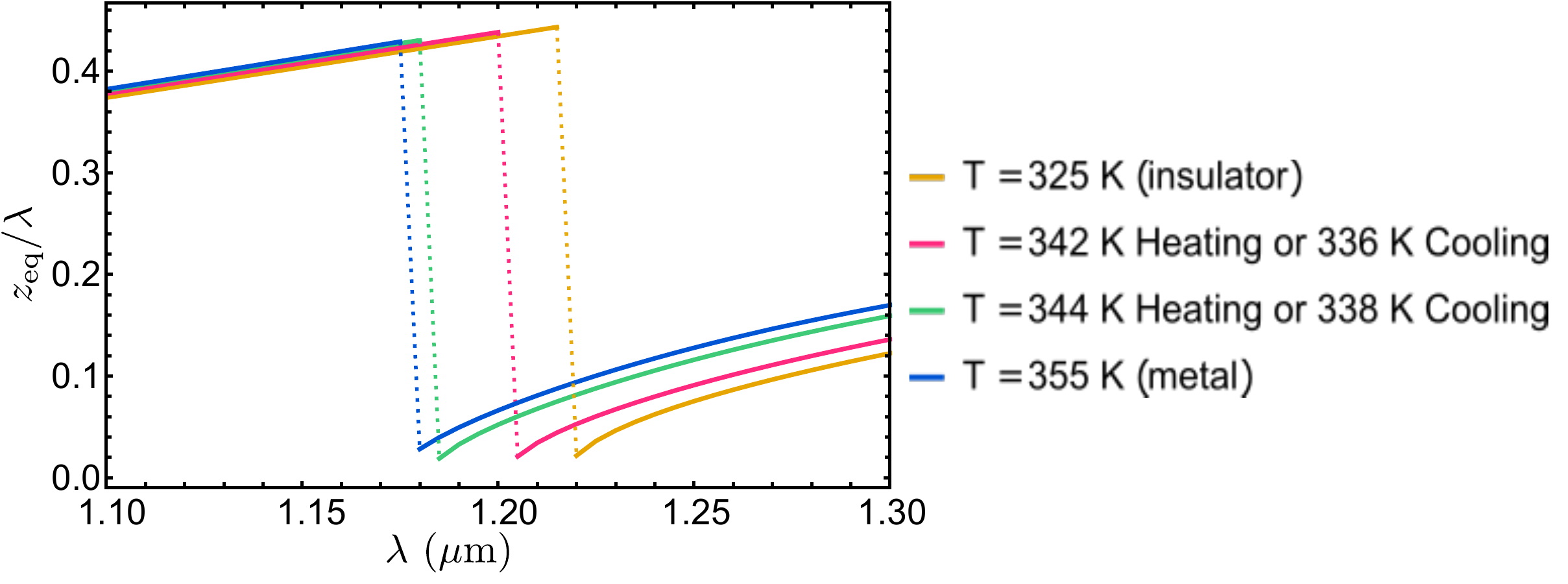}
	\caption{The normalized position $z_{\rm eq}/\lambda$ of the first equilibrium point as a function of $\lambda$ for different temperatures. Dotted lines indicate discontinuities in $z_{\rm eq} (\lambda)$ for each temperature.}
	\label{fig:StatPoint_ED}

\end{figure}

We also investigate the electric field distributions of the scattered field (Figure~\ref{fig:FieldDistributions_ED}) at a given instant of time. Surely, the field distribution is qualitatively altered when we compare insulator and metallic phases regardless the $\lambda$ considered. Recall that, at $z = 50$ nm, the force $F^e_\perp$ on the dipole changes from repulsive to attractive through heating for $\lambda = 1.2$ $\mu$m, but it is always attractive for $\lambda = 1.3$ $\mu$m [Figure~\ref{fig:FT_ED}(a)]. In terms of the field distributions, this fact can be inferred from the relative field orientations at the dipole position. The orientation changes from insulator to the metallic phase for $\lambda = 1.2$~$\mu$m, but it is essentially the same for $\lambda = 1.3$ $\mu$m. We stress that it is not the absolute orientation of the field that matters for unveiling the character of the force  - after all, the fields oscillate in time - but the fact that there is a relative orientations in the two phases that are flipped.

\begin{figure}[h!]

	\centering
	\includegraphics[width=0.98\linewidth]{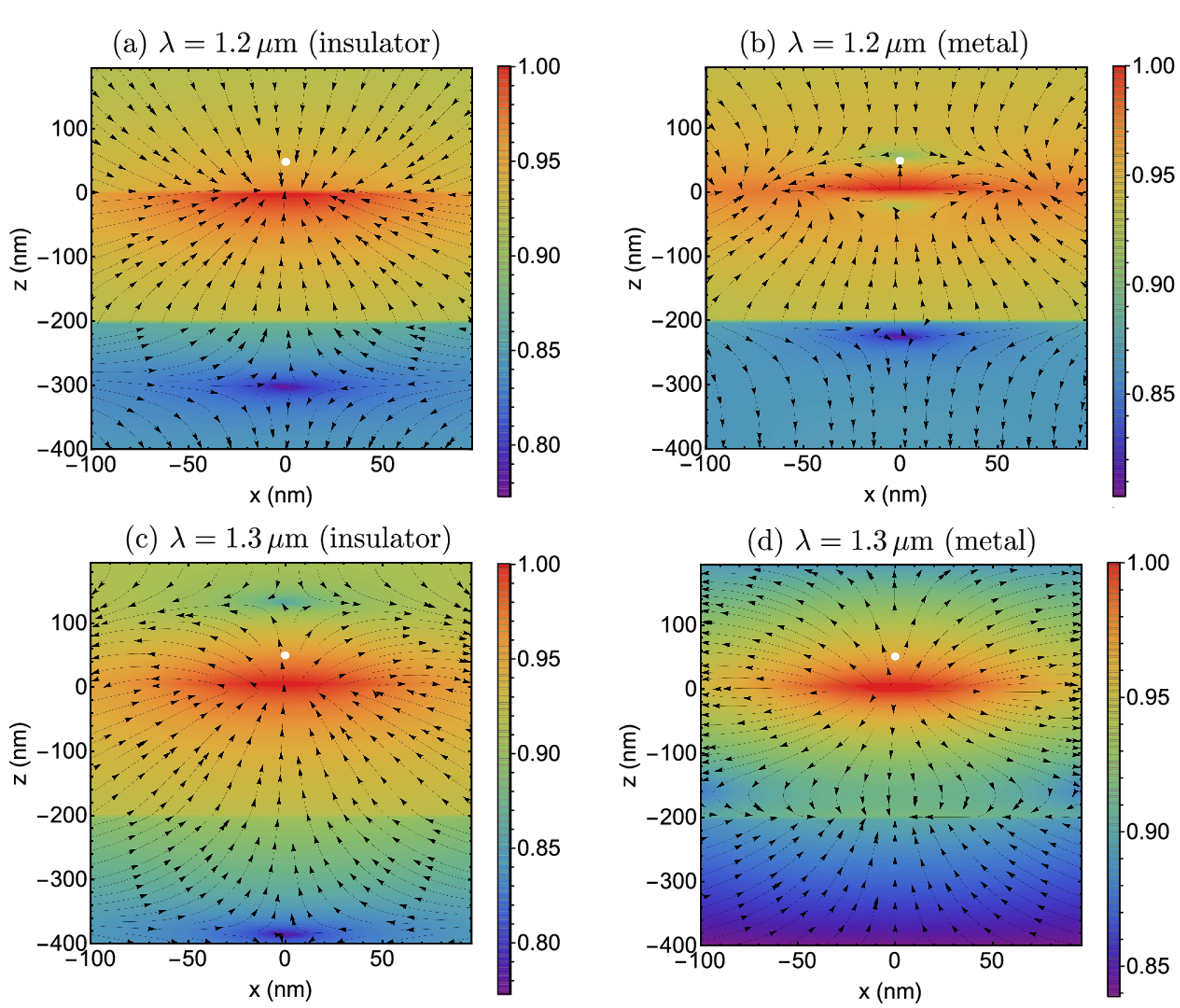}
	\caption{Scattered electric field distribution in the near-field regime for an oscillating electric dipole at $x = 0$ and $z = 50$ nm (white point in the plots) perpendicular to the VO$_2$ film in insulator [panels {\bf (a)} and {\bf (c)}] and metallic phases [panels {\bf (b)} and {\bf (d)}], considering the dipole wavelength $\lambda = 1.2$ $\mu$m [panels {\bf (a)} and {\bf (b)}] and $\lambda = 1.3$ $\mu$m [panels {\bf (c)} and {\bf (d)}]. The VO$_2$ film lies in the region $- 200$ nm $< z < 0$ nm and the sapphire substrate lies in the region $z< - 200$ nm. The electric field is normalized by its maximum value.}
	\label{fig:FieldDistributions_ED}

\end{figure}

It is also interesting to explore the changes of the electric field intensities exhibited in Figure \ref{fig:FieldDistributions_ED}. Recalling that the electric dipole is pushed to regions where the field intensities is greater, it can be realized that in the situations depicted in panels (b), (c) and (d), the dipole is attracted to the VO$_2$ film in agreement with Figure \ref{fig:FzLambdaT_ED}. Although in the situation described in panel (a) the dipole is repelled by the VO$_2$ film, this is not seen so clearly in this panel.

\begin{figure}[h!]

	\centering
	\includegraphics[width=0.42\linewidth]{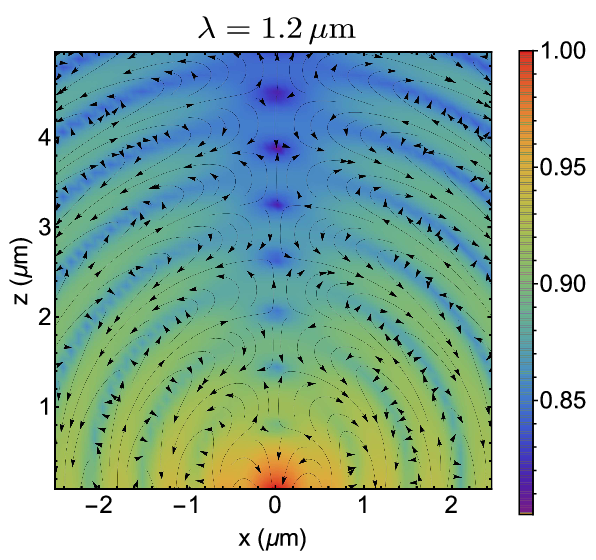}
	\caption{Scattered electric field distribution in the far-field regime for an oscillating electric dipole at $x = 0$ and $z = 50$ nm perpendicular to the VO$_2$ film in insulator phase for $\lambda = 1.2$ $\mu$m. The electric field is normalized by its maximum value in this region.}
	\label{fig:FieldDistributionsFF_ED}

\end{figure}

In Figure~\ref{fig:FieldDistributionsFF_ED} we calculated the scattered electric field distribution of an oscillating electric dipole perpendicular to the VO$_2$ film in the insulator phase with $\lambda = 1.2$ $\mu$m, but now, in the far-field regime. In this region, we checked that there is no qualitative changes between the distributions in metallic and insulator phases, as expected, since the more distant  the dipole is from the VO$_2$ film the less important are the boundary conditions and consequently, the less important are the difference between metals and insulators. Note the transverse character of the field and its oscillatory behavior with the distance, as expected. Observe that local minima are separated by a distance $\lambda/2$ and the field is attenuated for large distances.


\subsection{Magnetic Dipole}


Figure~\ref{fig:FT_MD} represents the magnetic force on the magnetic dipole as a function of temperature for different values of $\lambda$. Besides the thermal hysteresis, note that for $\lambda \sim 1$ $\mu$m the force behavior is significantly different for perpendicular and parallel configurations [Figures~\ref{fig:FT_MD}(a) and \ref{fig:FT_MD}(c)], but it is not for $\lambda \gtrsim 2.5$ $\mu$m [Figures~\ref{fig:FT_MD}(b) and \ref{fig:FT_MD}(d)]. Moreover, likewise the electric dipole case, there are wavelengths for which the force changes sign just by heating the VO$_2$ film. However, the wavelength interval for which this may occur is much larger here.

\begin{figure}[h]

	\centering
	\includegraphics[width=0.95\linewidth]{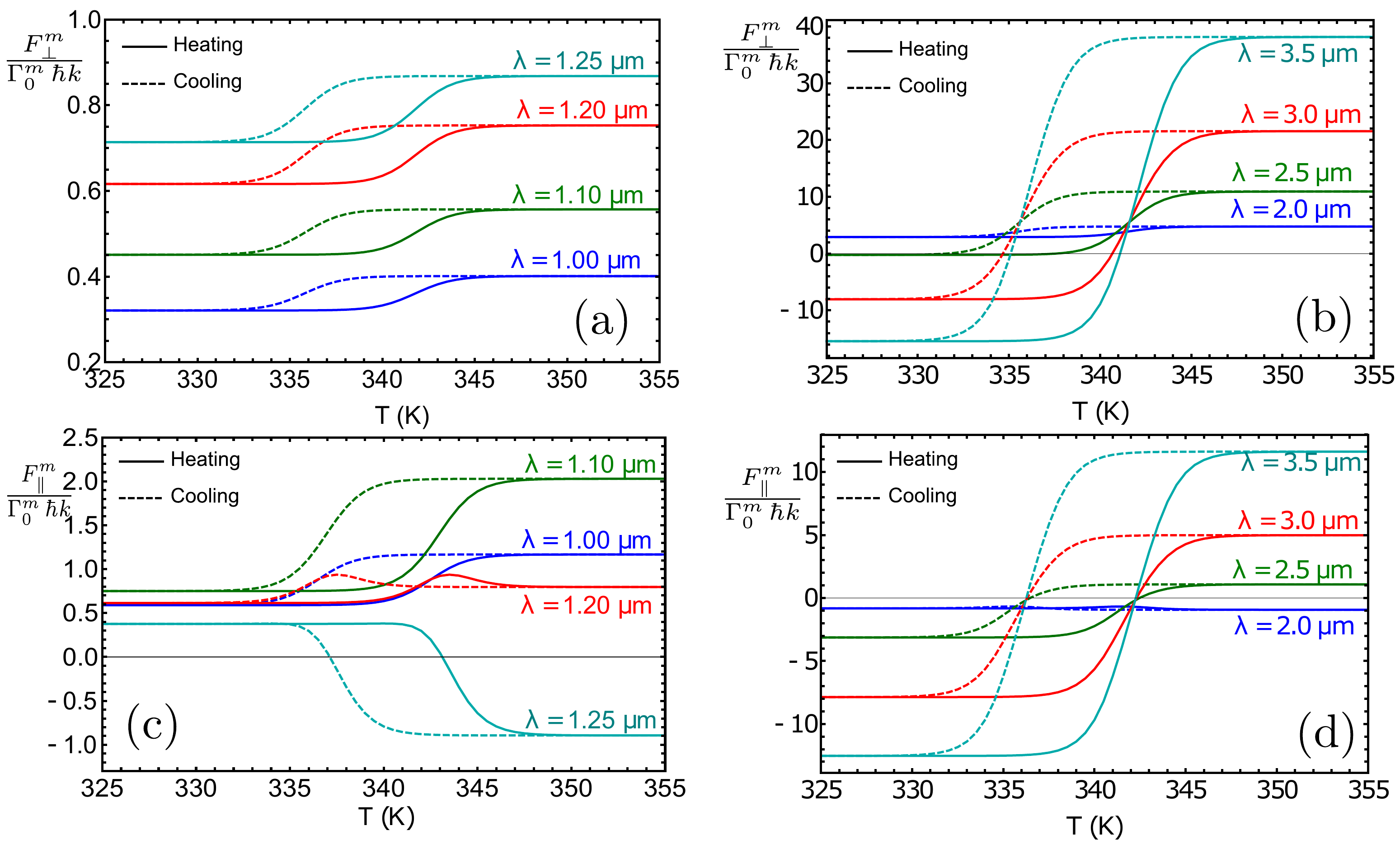}
	\caption{Force on an oscillating magnetic dipole as a function of temperature for different wavelengths in both perpendicular [panel {\bf (a)} and {\bf (b)}] and parallel configurations [panel {\bf (c)} and {\bf (d)}] for distinct wavelengths. The dipole is assumed to be at $z = 50$ nm from the VO$_2$ film.}
	\label{fig:FT_MD}

\end{figure}

Figure~\ref{fig:FLambda_MD} shows the magnetic force as a function of $\lambda$ in both parallel and perpedicular configurations. Comparing Figures~\ref{fig:FLambda_MD}(b) and \ref{fig:FLambda_MD}(c), one may note that, in the interval $1$ $\mu$m $< \lambda \lesssim 2.5$ $\mu$m, the forces for the perpendicular and parallel configurations of the dipole are qualitatively different. Particularly, the force may change from repulsive to attractive with heating in the interval $1.22$ $\mu$m $< \lambda < 1.31$ $\mu$m for a magnetic dipole parallel to the surface [see Figure~\ref{fig:FLambda_MD}(d)], but it is always repulsive in this range for a perpendicular configuration. Nevertheless, heating changes the force from attractive to repulsive in a much broader range than for the electric dipole case, i. e., $\lambda \gtrsim 2.5$ $\mu$m, regardless the dipole orientation (specifically for a parallel configuration this happens for $\lambda \gtrsim 2.3$ $\mu$m).

\begin{figure}[h]

	\centering
	\includegraphics[width=0.95\linewidth]{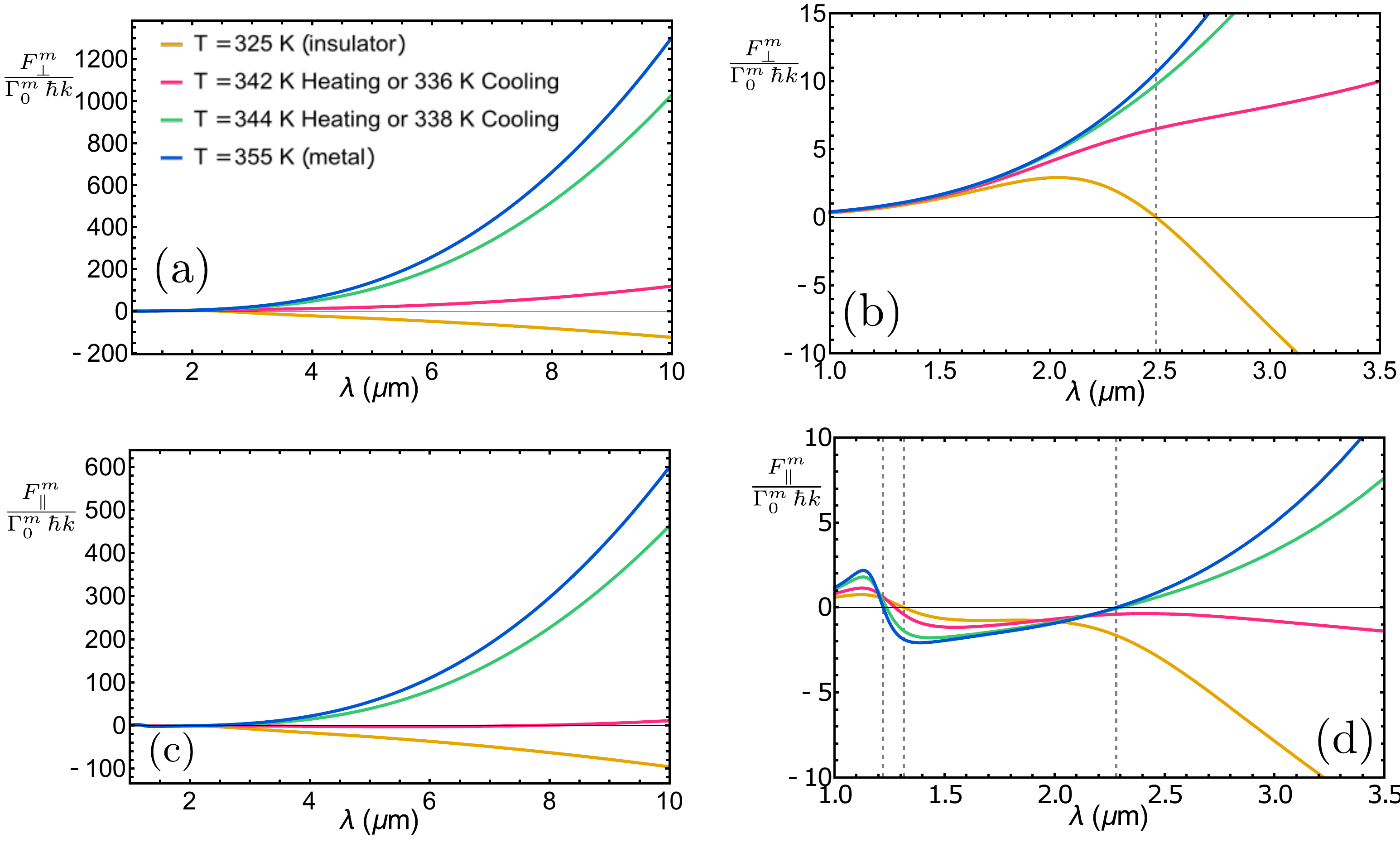}
	\caption{Force on an oscillating magnetic dipole as a function of $\lambda$ in both perpendicular [panels {\bf (a)} and {\bf (b)}] and parallel configurations [panels {\bf (c)} and {\bf (d)}] for distinct temperatures. Panels {\bf (b)} and {\bf (d)} show detailed intervals of panels {\bf (a)} and {\bf (c)}, respectively. The dipole is assumed to be at a distance $z = 50$ nm from the VO$_2$ film. Dashed lines indicate $\lambda$-intervals for which the force may change its sign depending on the temperature.}
	\label{fig:FLambda_MD}

\end{figure}

In order to further clarify these results, we also performed an extreme near-field ($c \rightarrow \infty$) approximation on Equations (\ref{eq:FzMPerp}) and (\ref{eq:FzMPar}), obtaining
\begin{align}
F^m_\perp &\simeq - \frac{\mu_0}{16 \pi} \, \frac{2 \, m_{z}^2 \, k^2}{4 z^2} \, \eta_\perp^m, && \eta_\perp^m = \frac{\mathcal{R}e \, \varepsilon_{\textrm{VO}_2}-1}{4} \,,    \label{eq:ForceNF_MD1} \\
F^m_\parallel &\simeq - \frac{\mu_0}{16 \pi} \, \frac{ m_{\parallel}^2 \, k^2}{4 z^2} \,   \eta_\parallel^m, && \eta_\parallel^m = \frac{\mathcal{R}e \, \varepsilon_{\textrm{VO}_2}-1}{4} + \frac{|\varepsilon_{\textrm{VO}_2}|^2 -1}{|\varepsilon_{\textrm{VO}_2}+1|^2} \,.
  \label{eq:ForceNF_MD2}
\end{align}
By comparison of these expressions with Equation~(\ref{eq:ForceNF_ED}), one can realize that, besides the factor 2 also present in the electric dipole case, the difference between parallel and perpendicular cases have now an additional term $\eta_\perp^m$. To clear up these differences, the factors $\eta^m_\perp$ e $\eta^m_\parallel$ are plotted in Figure~\ref{fig:ForcefactorMD}. Note that, for $\lambda~\gtrsim~2.5$~$\mu$m, the contribution $ \eta_e = (|\varepsilon_{\textrm{VO}_2}|^2 -1)/|\varepsilon_{\textrm{VO}_2}+1|^2$ only shifts up the force in the parallel configuration. This behavior is in agreement with the plot of $\eta^e$ in Figure~\ref{fig:ForcefactorED}(b). In fact, a comparison between Figures~\ref{fig:FT_MD}(b) and \ref{fig:FT_MD}(d) determines that, apart from a shift and a factor, the terms in parentheses in Equations~(\ref{eq:ForceNF_MD1}) and (\ref{eq:ForceNF_MD2}) have qualitatively similar behaviors. Despite that, for $1$ $\mu$m $< \lambda \lesssim 2.5$ $\mu$m, the effect of $\eta^e$ is non-negligible. This accounts for the differences in the predictions of parallel and perpendicular configurations in this interval [see Figures~\ref{fig:FT_MD}(a) and \ref{fig:FT_MD}(c)]. Specifically, $F_\parallel^m$ is non-monotonic with $\lambda$ in the metallic phase and, more importantly, it may change sign with heating in the interval $1.22$ $\mu$m $< \lambda < 1.31$ $\mu$m [Figure~\ref{fig:FT_MD}(c)], which does not happen for $F^m_\perp$ [Figure~\ref{fig:FT_MD}(a)].

\begin{figure}[h]

	\centering
	\includegraphics[width=0.95\linewidth]{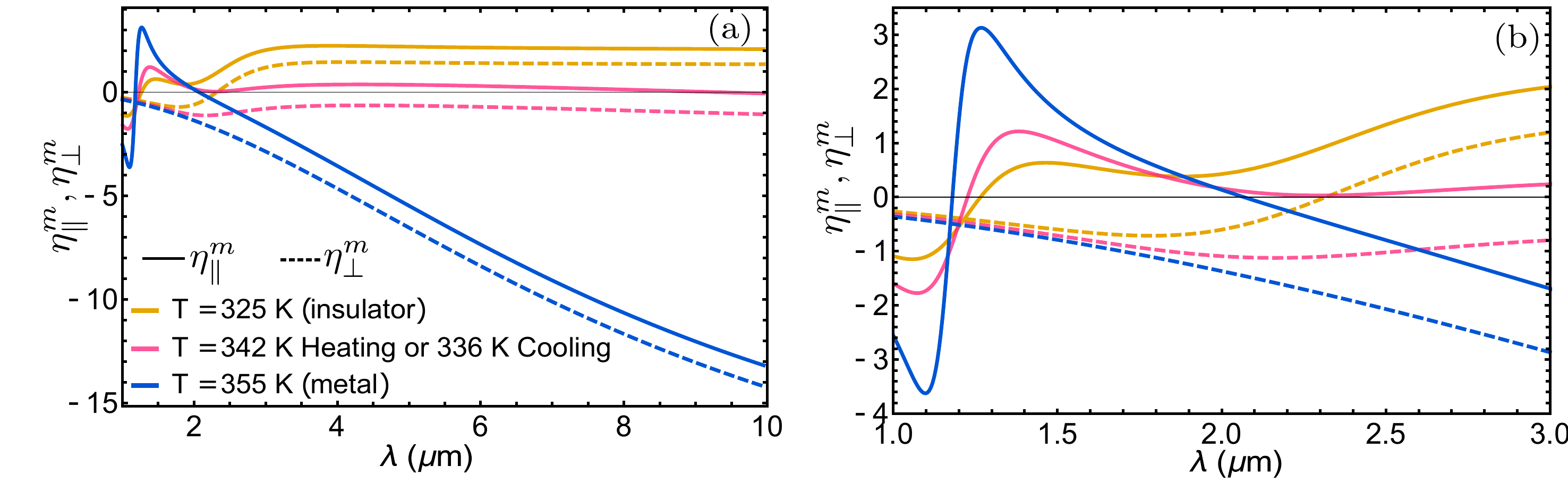}
	\caption{{\bf (a)} Factors $\eta_\perp^m$ (dashed lines) and $\eta_\parallel^m$ (solid lines) as functions of $\lambda$ for different temperatures. {\bf (b)} Details of the previous panel in the range $1$ $\mu$m $< \lambda < 3$ $\mu$m. }
	\label{fig:ForcefactorMD}

\end{figure}

The dependence of the force on the magnetic dipole with the term involving $\eta^m_\perp$ explains why the effect of changing the force sign is more robust in frequency for a magnetic dipole than for an electric dipole. Strictly speaking, the lower the frequency, the better is the distinction between metal and dielectric \cite{szilard2019hysteresis}. 
The distinction between the force in the metallic and insulator phases is clearer for a magnetic dipole than for an electric dipole since in the former the dependence on $\mathcal{R}e \, \varepsilon_{\textrm{VO}_2}$ is stronger [see Equations (\ref{eq:ForceNF_ED}), (\ref{eq:ForceNF_MD1}) and (\ref{eq:ForceNF_MD2})]. The only exception is found for higher frequencies where the distinction between metal and dielectric is blurred. Even though, in this region, we were still able to find a short $\lambda$-interval in which a change in the sign of the force on magnetic dipole occurs. However, the main reason for that is different, relying on the contribution $\eta^e$ with $\mathcal{R}e\, \varepsilon_{\textrm{VO}_2} \sim -1 $ and small $\mathcal{I}m\, \varepsilon_{\textrm{VO}_2}$.

In addition, Equations (\ref{eq:ForceNF_MD1}) and (\ref{eq:ForceNF_MD2}) reveal a scaling law with the distance in the form $z^{-2}$. Comparison with $z^{-4}$ scaling law for the electric case, given by Equation (\ref{eq:ForceNF_ED}), evidences a two powers difference between the electric and magnetic setups. It may be attributed to the fact that the electric field of an oscillating electric dipole ${\bm d} = {{\bm d}_0} \, e^{- i \omega t}$ has three terms ($1/r$, $1/r^2$ and $1/r^3$), whereas the electric field of an oscillating magnetic dipole ${\bm m} = {{\bm m}_0} \, e^{- i \omega t}$ has only two ($1/r$ and $1/r^2$)\cite{kort2011exact}. The reason for that is very simple: a static magnetic dipole does not create an electric field, so that there is no term proportional to $1/r^3$ in the expression of the electric field produced by an oscillating magnetic dipole. This fact also has interesting conseguences in the dispersion force between an electrically polarizable atom and a magnetically polarizable one, namely: while in the former case the non-retarded force between them is proportional to $1/r^7$ in the later it is proportional to $1/r^5$ \cite{Farina2002,feinberg1970general,farina2002force}. Likewise, the magnetic field of an oscillating magnetic dipole has three terms, while the magnetic field of an oscillating electric dipole has only two. In other words, the ultimate reason for such two powers difference is related to the fact that the magnetic field of an electric oscillating dipole does not contain the static-like term. Moreover, the scalling law $z^{-2}$ of the force on a magnetic dipole compared to $z^{-4}$ law of force on an electric dipole accounts for the main reason why the electric contribution is dominant in this near-field regime.

Figure~\ref{fig:FzLambdaT_MD} shows the force on a magnetic dipole as a function of $z$ for two distinct wavelengths, $\lambda = 2$ $\mu$m for which the force is repulsive in the near-field and $\lambda = 4$~$\mu$m for which the force changes sign in the near-field [recall Figure~\ref{fig:FT_MD}(b)]. Besides the expected oscillatory behavior for large distances, note that, for $\lambda = 2$ $\mu$m, the force $F_\perp^m$ diverges positively as $z\rightarrow 0 $ regardless the temperature while, for $\lambda = 4$ $\mu$m, there are temperatures for which the force diverges negatively as $z \rightarrow 0$. In fact, the latter behavior is present for $\lambda \gtrsim 2.5$ $\mu$m. In contrast with the electric case, there are no wavelengths for which the force on the magnetic dipole diverges only negatively as the distance $z$ decreases for different temperatures.

\begin{figure}[h]

	\centering
	\includegraphics[width=0.95\linewidth]{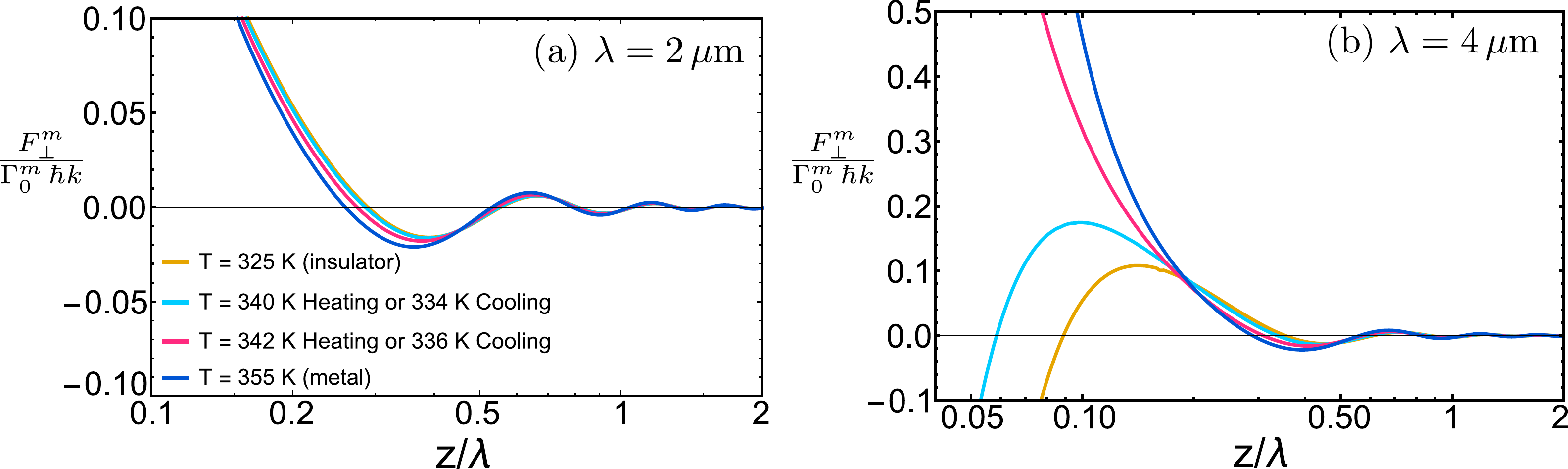}
	\caption{Force on the magnetic dipole $F_\perp^m$ as a function of distance $z$, for different temperatures and for {\bf (a)} $\lambda = 2$ $\mu$m and {\bf (b)} $\lambda = 4$ $\mu$m.}
	\label{fig:FzLambdaT_MD}

\end{figure}

\begin{figure}[h]

	\centering
	\includegraphics[width=0.9\linewidth]{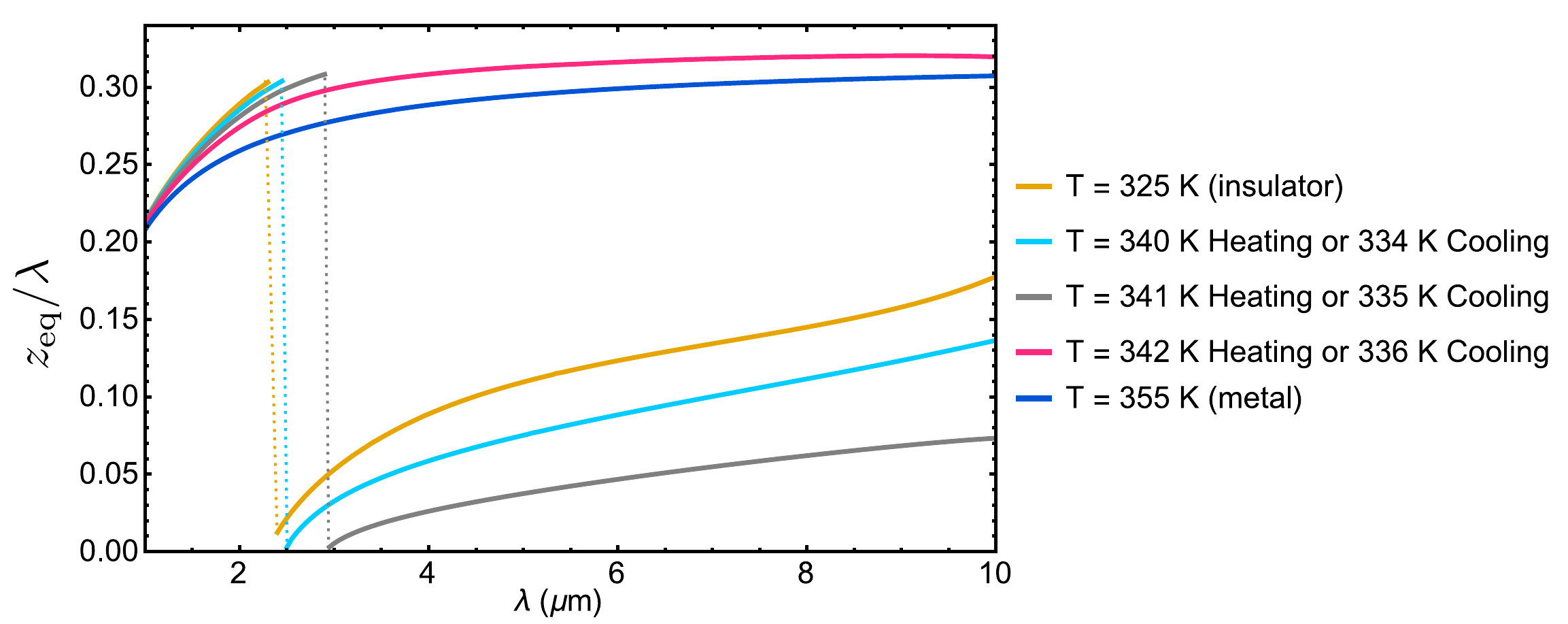}
	\caption{The normalized position $z_{\rm eq}/\lambda$ of the first equilibrium point of the magnetic dipole and VO$_2$ film system as a function of $\lambda$ for different temperatures. Dotted lines indicate discontinuities in $z_{\rm eq} (\lambda)$ for each temperature. Note that, for $T > 342$ K, there are no discontinuities in the range considered and the first equilibrium point is always unstable.}
	\label{fig:StatPoint_MD}

\end{figure}

These conclusions about whether the force diverges positively or negatively as $z \rightarrow 0$ can also be driven from Figure~\ref{fig:StatPoint_MD}, that shows the position of the first equilibrium point as a function of the wavelength $\lambda$. Similarly to the electric case, the discontinuity on the plot informs about the stability of the equilibrium point. For $\lambda < 2.3$ $\mu$m, the first equilibrium point is always unstable. As $\lambda$ increases, its stability depends on the temperature. Interestingly, for $T > 342$ K, it is always unstable. This result is in agreement with the prediction that there is a $\lambda$ above which the force on the magnetic dipole in the perpendicular configuration always changes its sign with heating, i. e., $\lambda \sim 2.3$ $\mu$m. From the previous discussion, we can verify that, in order to have the possibility of changing the attractive/repulsive character of the force by varying the temperature, the position of the dipole and its oscillating frequency need to be properly chosen.

\begin{figure}[h!]

	\centering
	\includegraphics[width=0.98\linewidth]{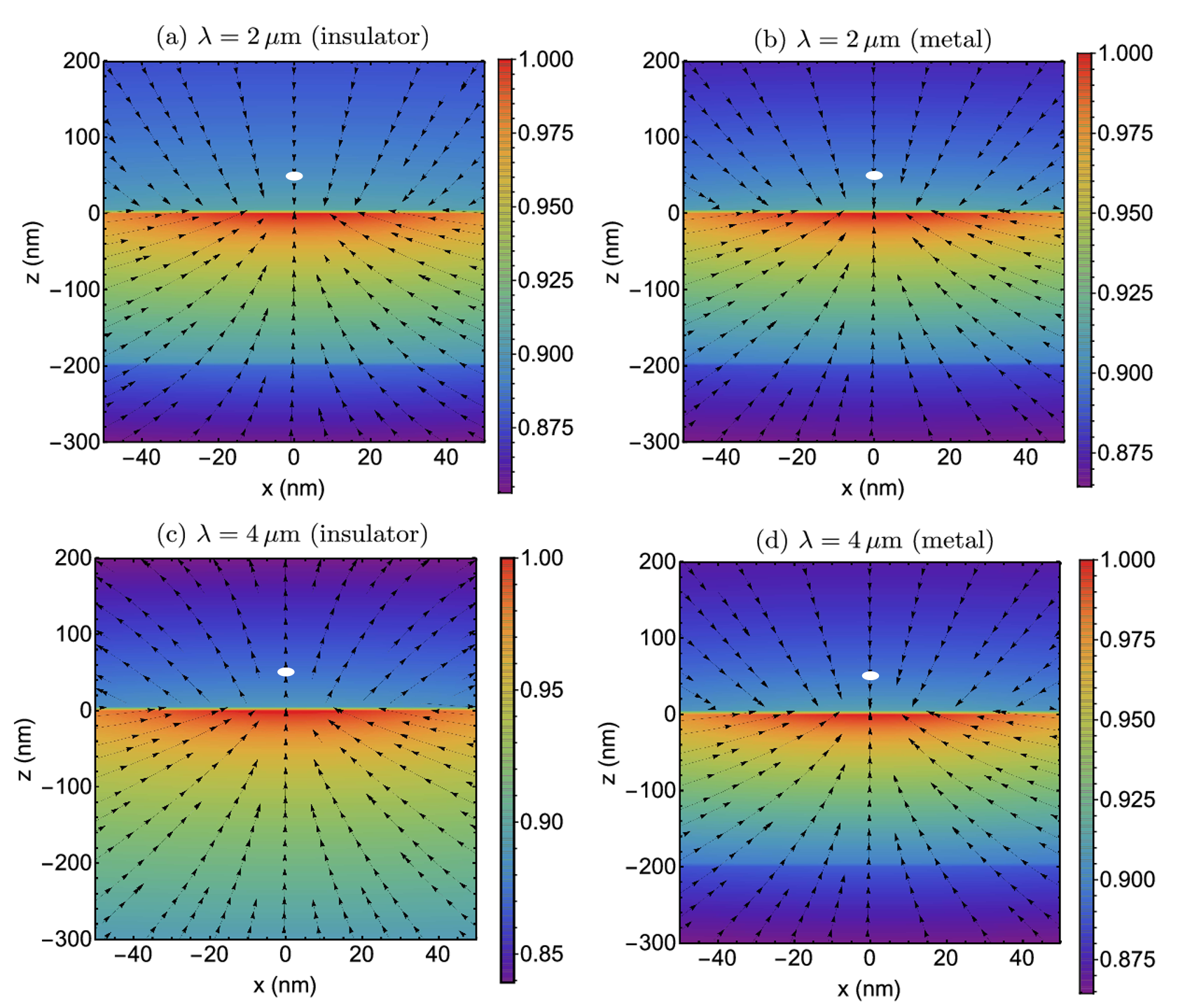}
	\caption{Scattered magnetic field distribution in the near-field for an oscillating magnetic dipole at  $x = 0$ and $z = 50$ nm (white point in the plots) perpendicular to the VO$_2$ film in insulator [panels {\bf (a)} and {\bf (c)}] and metallic phases [panels {\bf (b)} and {\bf (d)}], considering the dipole wavelength $\lambda = 2$ $\mu$m [panels {\bf (a)} and {\bf (b)}] and $\lambda = 4$ $\mu$m [panels {\bf (c)} and {\bf (d)}]. The VO$_2$ film lies in the region $-200$ nm $< z < 0$ nm and the sapphire substrate lies in the region $z< - 200$ nm. The magnetic field is normalized by its maximum value in this region.}
	\label{fig:FieldDistributions_MD}

\end{figure}

Lastly, we computed in Figure~\ref{fig:FieldDistributions_MD} the magnetic field distributions for a perpendicular magnetic dipole at a given instant of time. Our considerations are very similar to the ones made before about the electric field distributions for an electric dipole (Figure~\ref{fig:FieldDistributions_ED}). When the force does not change its sign, the field at the dipole position also does not alter its orientation, as occurs, for example, for $\lambda = 2$ $\mu$m. In contrast, when the force changes from attraction to repulsion with heating, it can be seen through the modification on the field orientation on the particle position (for $\lambda \gtrsim 2.5$ $\mu$m). In particular, for $\lambda = 2$~$\mu$m, the magnetic field at the dipole position points downwards in both insulator and metallic phases [Figures~\ref{fig:FieldDistributions_MD}(a) and \ref{fig:FieldDistributions_MD}(b)] while, for $\lambda = 4$ $\mu$m, it points upwards in the insulator phase [Figure~\ref{fig:FieldDistributions_MD}(c)] and downwards in the metallic phase [Figure~\ref{fig:FieldDistributions_MD}(d)]. Regarding the analysis of the field intensities, we should point out that, differently from the electric case, it is not possible to discern the sign of the force as the gradients are too smooth within the panel's resolution.

In Figure~\ref{fig:FieldDistributionsFF_MD} we calculated the magnetic field distribution in the far-field regime (compare with Figure~\ref{fig:FieldDistributionsFF_ED}). In this region, there is no qualitative changes between the behavior of the distributions in metallic and insulator phases. Note that the separation between two local minima is again given by $\lambda/2$, the field is weakened and oscillates with the distance.

\begin{figure}[h!]

	\centering
	\includegraphics[width=0.42\linewidth]{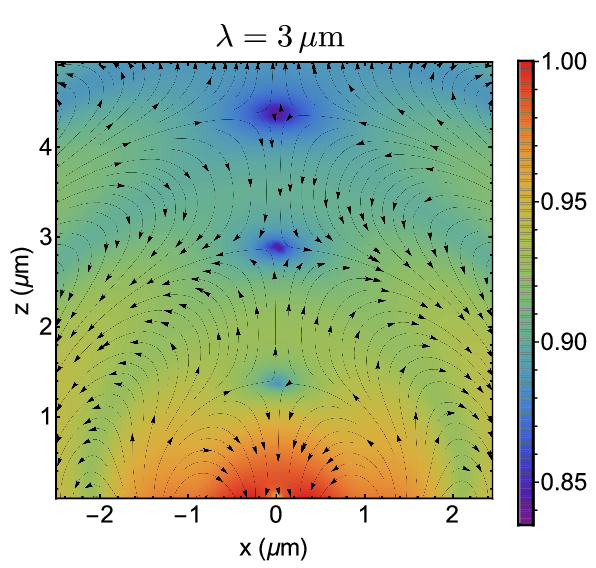}
	\caption{Scattered magnetic field distribution in the far-field regime for an oscillating magnetic dipole at $x = 0$ and $z = 50$ nm perpendicular to the VO$_2$ film in insulator phase for $\lambda = 3$ $\mu$m. The magnetic field is normalized by its maximum value in this region.}
	\label{fig:FieldDistributionsFF_MD}

\end{figure}


\section{Final remarks and conclusions}
\label{SecConclusions}



In this work, we investigated optical forces acting on oscillating electric and magnetic dipoles close to a VO$_2$ phase-change film, in order to explore the effects of its well-known metal-insulator transition and low thermal hysteresis. Additionally, we considered two distinct configurations of the dipoles in our setup, to wit, perpendicular and parallel to the film. We verified the remarkable possibility of thermally controlling the optical force on both dipoles in the near-field regime, once we choose appropriately the dipole frequency and its distance from the VO$_2$ surface. More specifically, we show that the thermal hysteresis allows for a change in the attractive/repulsive character of the force just by heating (or cooling) the VO$_2$ film. Interestingly, the force on electric and magnetic dipoles shows different power laws with the distance from the film. Furthermore, for magnetic dipoles, there are more wavelength intervals for which the force may change its sign depending on the temperature. Altogether, we hope that our results further expand the broad spectrum of applications of these materials, providing alternative ways to tune light-matter interactions using phase-change materials.


\vspace{6pt} 



\authorcontributions{
D.S. and W.K.-K. conducted the numerical calculations. All authors analyzed the results and contributed to this work. All authors have read and agreed to the published version of the manuscript.}

\funding{C.F. and F.S.S.R. acknowledge Conselho Nacional de Desenvolvimento Científico e Tecnológico (CNPq) for financial support (grant numbers 310365/2018-0 9 and 309622/2018-2). F.S.S.R. (grant number E26/203.300/2017) and P.P.A. acknowledge Fundacão de Amparo à Pesquisa do Estado do Rio de Janeiro (FAPERJ). D.S. and F.A.P. also acknowledge the funding agencies. W.K.-K. acknowledges the Laboratory Directed Research and Development program of Los Alamos National Laboratory for funding under Project No. 20210327ER.}


\conflictsofinterest{The authors declare no conflict of interest.}

\abbreviations{The following abbreviations are used in this manuscript:\\

\noindent 
\begin{tabular}{@{}ll}
MIT & metal–insulator transition\\
BEMT & Bruggeman effective medium theory\\
\end{tabular}}

\end{paracol}
\reftitle{References}

\externalbibliography{yes}
\bibliography{bibliography}

\end{document}